\documentstyle[aps]{revtex}

\begin{document} 
\title{Higgs boson production via Double Pomeron Exchange at
the LHC}
 \author{M.  Boonekamp\thanks{  CERN, CH-1211, Geneva 23,
Switzerland and CEA/DSM/DAPNIA/SPP, CE-Saclay, F-91191
Gif-sur-Yvette Cedex, France}, A.  De Roeck\thanks{ CERN, CH-1211, 
Geneva 23,
Switzerland}, R.  Peschanski\thanks{ CEA/DSM/SPhT, Unit\'e de 
recherche 
associ\'ee au CNRS, CE-Saclay, F-91191
Gif-sur-Yvette Cedex, France} and C.  Royon\thanks{ 
CEA/DSM/DAPNIA/SPP, 
CE-Saclay, F-91191
Gif-sur-Yvette Cedex, France
 and Texas U. at Arlington, USA}}
\maketitle

\begin{abstract} 
We study  Higgs boson production via Double Pomeron Exchange 
allowing for the presence of Pomeron remnants.  We estimate the 
number of events produced at the LHC collider, as a function of the 
Higgs boson mass and its decay channel. The model which 
successfully describes the high mass dijet spectrum observed at 
Tevatron (run I) is used to  predict rates of events with tagged
protons for the acceptance range of the CMS/Totem experiments.  
Sizeable cross-sections and encouraging event selection signals are 
found, demonstrating especially for   smaller Higgs boson masses
the importance to study the diffractive channels. Tagging of the 
Pomeron remnants can be exploited  to achieve  a good resolution on 
the Higgs boson mass for inclusive diffractive events, by 
optimizing  an intermediary  analysis between  higher 
cross-sections of the fully {\it inclusive} mode (all Pomeron 
remnants) and cleaner signals of the {\it exclusive} mode (without 
Pomeron remnants).
\end{abstract}

 \bigskip 
 
 {\bf 1. Introduction.}  It has been suggested \cite{bi90,bi91} 
that
diffractive
Higgs boson production via Double Pomeron Exchange (DPE) is an 
interesting
channel to study the Higgs boson at hadron colliders.  
The lack of a solid QCD based  framework for
diffraction made a purely theoretical study difficult. Recently
\cite{us} however, the possibility of a better determination of the
cross-sections and event rates was proposed, using a model allowing 
the joint
description of Higgs boson production and of the observed high mass 
dijet
production at the 
Tevatron (run I) \cite{CDF0}, allowing to compare and 
normalize the predictions
to the  data using a (simplified) simulation of the
detector.  
One important difference with previous (purely {\it 
exclusive}) estimates lies on the
consideration of {\it inclusive }  production
 $pp \rightarrow p +X +H +Y +p$, see Fig.~1, namely
with particles accompanying the Higgs/dijet production in the 
central region. We 
will call $X,Y$ the ``Pomeron remnants'' in the following. 
The presence of the Pomeron remnants is vital for the 
good description of the dijet mass spectrum as discussed 
in~\cite{us}.

Further estimates
using a different Pomeron model \cite {allinclu}-a gives different 
numbers,
since important sources of uncertainties still remain, 
but confirm the viability
of such processes for the LHC.  Pioneering studies which will be 
possible
in the near future at the
Tevatron \cite{us,al00}, on the evaluation of experimental 
possibilities using
outgoing (anti)proton tagging, on the resolution
achievable with missing mass methods and with information on the
Pomeron remnants, can be used to pin down remaining
theoretical uncertainties. It can  upgrade diffractive production 
to a complementary tool for the
analysis of the Higgs boson characteristics at the LHC.

Our aim is to give predictions for  Higgs boson production at the 
LHC based on
{\it inclusive}  dijet
production at colliders via DPE.  Thus, we {\bf i)} normalize the 
theoretical predictions to the observed
dijet rate at the Tevatron ($p\overline{p}$ collisions with 
$\sqrt{s}= 1.8$ TeV), and assume the obtained normalization 
factor is also valid at the LHC ($pp$ collisions at $\sqrt{s}=14$ 
TeV),
{\bf
ii)} obtain a 
prediction for the Higgs boson production cross-sections and event 
rates at the LHC and, 
{\bf iii)} discuss how the experimental opportunities  at the  LHC 
can 
be used for precision measurements.

In section {\bf 2}, we use as a starting point our model \cite{us}  
based on 
the extension  to {\it 
inclusive} diffractive production of the Bialas-Landshoff {\it 
exclusive}
model for Higgs boson and heavy flavor jet production \cite{bi90}. 
We showed 
\cite{us} that this is  able to reproduce the observed 
distributions, in 
particular the dijet
mass fraction spectrum, the normalization being fixed from 
experiment. 

The main lesson of our study is that an interesting  potential for 
the
Higgs boson studies at the LHC can be expected in double
proton tagged experiments.  In section {\bf 3} , we derive the 
predicted  number
of events 
as a function of $M_H$, depending on experimental cuts and 
the Higgs decay channel, and in section {\bf 4} the potentialities 
for Higgs 
mass 
reconstruction.

{\bf 2. Formulation.}  Let us introduce the formulae for {\it 
inclusive} Higgs boson
and dijet production cross-sections via DPE \cite{us}: 
\begin{eqnarray}
d\sigma_H^{incl} &=& C_{H}\left(\frac {x^g_1x^g_2 s
}{M_{H}^2}\right)^{2\epsilon}\ \delta \left(\xi _1 \xi
_2\!-\!\frac{M_{H}^2}{x^g_1x^g_2 s} \right) \ \prod _{i=1,2}
\left\{G_P(x^g_i,\mu)\ dx^g_i  d^2v_i\ \frac {d\xi _i}{1\!-\!\xi 
_i}\
{\xi _i}^{\alpha' v_i^2}\ \exp \left(-2 
v_i^2\lambda_H\right)\right\}, \nonumber
\\ 
d\sigma_{JJ}^{incl} &=& C_{JJ}\left(\frac {x^g_1x^g_2 s }{M_{J
J}^2}\right)^{2\epsilon}\!\!  F_{JJ}\ \delta ^{(2)} \left(\sum 
_{i=1,2}
(v_i\!+\!k_i)\right) \prod _{i=1,2} \left\{G_P(x^g_i,\mu)\  {d\xi 
_i} \ {d\eta 
_i}\ 
d^2v_i\ d^2k_i\ {\xi _i}^{\alpha' v_i^2}\
\exp \left(-2 v_i^2\lambda_{JJ}\right)\right\} \ , 
\label{dinclu}
\end{eqnarray}
where $x^g_1, x^g_2$ define, on each side (see Fig.1),  the 
fraction of the 
Pomeron's 
momentum carried by
the gluons involved in the hard process and $G_P(x^g_{1,2},\mu),$
is, up to a normalization, the gluon structure function in the 
Pomeron
extracted \cite{ba00} from HERA experiments;  $\mu^2$ is the hard 
scale
(for simplicity kept fixed at $75 $GeV$^2,$ the highest value 
studied at 
HERA; we
neglect the small \cite{ba00} contribution of quark initiated 
processes in the
Pomeron);  $\eta_1,\eta_2$ are the 
rapidities of the two jets and are defined as 
a function of the other kinematical variables by 
\begin{equation}
m_{T1}^{f}e^{\eta_1}+m_{T2}^{f}e^{\eta_2}=\xi_1 x^g_1\ ;\
m_{T1}^{f}e^{-\eta_1}+m_{T2}^{f}e^{-\eta_2}=\xi_2 x^g_2
\ , 
\label{rapidities}
\end{equation}
where $m_{Ti}^{f}$ are the transverse mass of the quark with flavor 
$f.$

 The formulae (\ref{dinclu}) are written for a Higgs boson of mass 
$M_H$ and two
jets (of total mass $M_{JJ}$), respectively.  The Pomeron 
trajectory is
$\alpha (t)=1+\epsilon+\alpha't$ ($\epsilon\!\sim 
\!.08,\alpha'\!\sim \!.25 \
GeV^{-2}$), $\xi _{1,2}\ (<0.1)$ are the Pomeron's fraction of 
longitudinal
momentum, $v_{1,2},$ the 2-transverse momenta of the outgoing $p 
\bar p$,
$k_{1,2}$ those of the outgoing quark jets, $\lambda_H \sim 2\ 
GeV^{-2}$ (resp. 
$\lambda_{JJ} \sim 3\ GeV^{-2}$) the
slope of the  non perturbative  coupling for the Higgs boson (resp. 
dijets), and 
the 
constants $C_H , C_{JJ}$
are normalizations  including  a non-pertubative
gluon coupling \cite{bi90}, appreciably cancelled  in the ratio 
$C_H / C_{JJ}.$

The dijet cross-section $\sigma_{JJ}$ depends on the $gg\to \bar 
Q^{f} Q^{f}$
 and $gg\to gg$ cross-sections \cite{co83}.  This gives for five 
quark flavors
  \begin{eqnarray}
   & & F_{JJ}= \sum _{f} F_{\bar Q^{f}Q^{f}}
 \left(\rho^{f}\right) + {54} \ F_{gg}\left(\rho^{gg}\right)\ ; \ 
\rho^{f}\equiv
 \ \frac {4\ m_{T1}^{f}m_{T2}^{f}}{M_{\bar Q^{f} Q^{f}}^2} ;\ 
\rho^{gg}\equiv \
 \frac {4\ p_{T1}p_{T2}}{M_{gg}^2}\ ,\nonumber\\ & & F_{\bar 
Q^{f}Q^{f}} \equiv
 \frac { \rho^{f}} {m_{T1}^{2\ f}m_{T2}^{2\ f}}\ \left(1-\frac
 {\rho^{f}}2\right)\left(1-\frac {9\ \rho^{f}}{16}\right) \ ;\ 
F_{gg} \equiv \
 \frac 1 {p_{T1}^2p_{T2}^2}\ \left(1-\frac {\rho^{gg}}4\right)^3\ .
 \label{rhoi}
  \end{eqnarray}
The colour factor  ${54}$ appears in  the ratio of
gluon jets {\it vs.} quark jets partonic cross-sections 
\cite{co83}.

 Note that the $gg\to \bar Q^{f} Q^{f}$ cross section depend on 
{\it transverse}
and not on {\it rest} quark masses.  Thus, all 5 quark flavors 
sizeably 
contribute to
the dijet cross-section.  This is to be contrasted with the {\it 
exclusive} case 
which  is
proportional to {\it rest}  masses \cite{bi90,us}, and thus 
considerably smaller.  

The physical origin of formulae (\ref{dinclu}) is the following:  
since the 
overall partonic configuration is
produced initially by the long-range, 
soft DPE interaction, we assume that, up to
a normalization, the {\it inclusive} cross-section is the 
convolution of the
``hard'' $partons \to Higgs \ boson,$ (or $partons \to jets$) 
subprocesses by 
the
Pomeron structure function into gluons, see Fig.1.  The expected 
factorization
breaking of hadroproduction will appear in the normalization 
through a
renormalization of the Pomeron fluxes, which are not the same as in 
hard
diffraction at HERA.  Indeed, this ansatz remarkably reproduces the 
dijet mass
fraction seen in experiment, see Ref.\cite{us}. This model has been 
successfully 
applied 
\cite {us} to Tevatron data on dijets and to predictions for Higgs 
production 
possibilities at the same accelerator 
\cite {us}.

{\bf 3. Predictions.}  
We can now give predictions for the Higgs boson production cross 
sections in
DPE events at the LHC, i.e. with  $\xi_{1,2} < 0.1$, 
by scaling our results by the same factor used for
 the CDF measurement on dijet cross sections, which means
increasing it  by
a factor 3.8 \cite {us}. 
The results are given in Table 1 (first column) and in Fig. 2.  
We note the high values of the cross-sections. Since the typical 
luminosity for
the LHC will be of the order of 10-100 fb$^{-1}$/year, this leads 
to several
thousand Higgs particles diffractively produced per year, even at 
low
luminosity. It is of course imperative
to study the corresponding background, on which we will comment 
after final selection in section {\bf 4}.
Hence for the inclusive channel the cross section is large, much 
larger
than recent calculations for the exclusive channel (see  
Ref.\cite{allinclu}-b).
In Ref.\cite{allinclu}-a,  similar conclusions are reached (note 
however 
\cite{allinclu}-c).

In Fig. 2, we show the effects of the 
 acceptance of the possible roman pots detectors at the LHC, which 
are
used in conjunction with a central detector. Following 
 ideas presently  discussed in a common study group
of the central detector CMS~\cite{cms} and the elastic/soft 
diffraction
experiment TOTEM~\cite{totem}, which both will use the 
same interaction region at the LHC, we 
choose four possible
configurations for roman pot detectors to measure the scattered 
protons.
The used acceptance numbers are based on~\cite{orava}.
 The first one (see, {\it Config.1} on   
Fig.2)
has  roman pot detectors located in the warm region of the LHC
respectively at 140-180 meters and 240 meters and assumes
a good acceptance for protons with $|t| < 2$ GeV$^2$, and 
$\xi>0.01$.
The second one ({\it Config.2}) considers only roman pots at 
140-180 meters 
\cite{totem}
and gives a good acceptance only for $|t| < 1.5$ GeV$^2$, and 
$\xi>0.02$.
{\it Config.3} assumes the presence of roman pot detectors 
in the cold region of the LHC at about 425 meters and gives a good
acceptance for $|t| < 2$ GeV$^2$, and $0.002 < \xi <0.02$. 
Certainly the latter will be challenging both from the machine and 
experimental point of view, but
Fig.2 demonstrates that 
such  detectors are needed to obtain a good acceptance for low mass
Higgs production.
{\it Config.4} corresponds to the full system
using all detectors.
In Table 1, the acceptance of the roman pot detectors in the
case of {\it Config.4} is taken into account, and we give the 
number of events
for 10 fb$^{-1}$ in the different Higgs decay modes.

{\bf 4. Higgs boson mass reconstruction.} The advantage of DPE 
events with respect to standard
Higgs boson production lies in offering a potential
to  reconstruct the Higgs particle parameters more precisely. For
example, one can hope to obtain
 a very precise Higgs mass reconstruction if one can tag and 
measure
both the protons in the roman pot detectors as well as
 the Pomeron remnants. 
 
 In Fig. 3
(upper part), we show the
distribution in pseudo-rapidity of the tagged proton
(highest $| \eta |$), the Pomeron remnants at the parton level
(medium $| \eta | 
$) and the Higgs
decay products for a Higgs mass of 120 GeV in our model. In Fig. 3 
we also give 
the 
size
of the rapidity gap between the tagged proton and the Pomeron 
remnant in the
case of a Higgs mass of 120 GeV (middle left), 700 GeV (middle 
right). 
In the last row of this figure we show the distance between
the Pomeron remnant and the nearest jet from the Higgs,
 for the two Higgs masses. The rapidity
distance  is large at the parton level. This region will be however 
mostly filled with soft particles from QCD radiation and from 
hadronization of 
the  
partons. We nevertheless expect  that the
remnants will remain visible after a cut on soft 
activity\footnote{Such a cut 
may 
anyway be required in the analysis to 
reduce the effects of soft inelastic overlaid pile-up  events due 
to 
multiple interactions per bunch crossing at the LHC.} (e.g. 
particles
below 1 GeV).

In Fig. 4, we describe the results of the
Higgs boson mass reconstruction assuming
 one is able to select and measure  the Pomeron
remnants. Then the 
 Higgs boson mass can be reconstructed by applying  quadri-momentum 
conservation to all particles in the final state, namely 
the Higgs boson, the scattered protons in the roman pot detectors, 
and the 
Pomeron
remnants. The energy $E = \sqrt{ \xi_{1} \xi_{2} s}$ is used to 
produce
the Higgs boson and the Pomeron remnants. Detecting these remnants
requires the presence of  detectors with an 
acceptance at high pseudo-rapidity, 
ideally up to $ |\eta| \ =10$ (see Fig.~3), but already
taggers up to a rapidity of 7.5-8 \cite{cms} give a good 
acceptance for a Higgs boson with a mass of 120 GeV. 
With such  taggers,
it is possible to reconstruct Higgs boson masses\footnote{The
precision obtained on the Higgs boson mass reconstruction using 
standard events
is quite high for high Higgs boson masses. Our method is especially 
useful at 
low 
Higgs
masses where the measurement is harder for standard events.} up to
 about 600 GeV.   

The quality of the reconstruction of the remnants needs to be 
demonstrated after including
hadronization and detector effects, and are subject to a future 
detailed study.
Meanwhile we assume it can be 
done with a  resolution of 
$100 \% / \sqrt E$ 
(resp. $300 \% / \sqrt E$ ) in an optimistic (resp. more 
pessimistic) scenario.
The mass distribution is then determined from the missing mass 
measurement
from the scattered protons, and the remnants are subtracted. 
The smearing on the resulting  Higgs
mass distribution is mainly due to the Pomeron
remnant measurements, hence a good resolution can be obtained when 
the
energy of the Pomeron remnants is small, {\it i.e.} a configuration 
nearer to 
the {\it 
exclusive} case.

In Fig. 4, we display the resolution (resp. 2.1, 4.0, 4.6 and
6.6 GeV) on the Higgs boson mass reconstruction
for four different cuts on the Pomeron remnant energy (resp. 20, 
50, 100
and 500 GeV) and for the optimistic 
scenario. This does not take into account the additional resolution 
smearing of 1-2 GeV expected from the missing mass analysis of the 
scattered protons~\cite{orava}.
The plot with the best resolution (2.1 GeV) is shown for
a  luminosity of 30 fb$^{-1}$. 
When the configuration is close to  an {\it exclusive} 
process, {\it i.e.} the remnant energy is less than 20 GeV,
this leads respectively to a resolution of 
2.1, 3.6
and 4.7 GeV for a Higgs boson mass of 120, 200 and 500 GeV.
In the more pessimistic scenario,  the resulting Higgs mass 
resolution is about 
7 
GeV for 
a 120 GeV Higgs.
A good coverage 
in pseudo-rapidity will be essential to be able to
 precisely measure  the Higgs
boson
parameters. Note that the events showing little energy for the 
Pomeron
remnants have a low value of $\xi$, since
$M_{H} \sim \sqrt
{ \xi_{1} \xi_{2} s}$, which leads to $\xi_{1} \xi_{2} \sim 7. 
10^{-5}$.
Hence roman pot detectors in the region of about 400 m from the 
interaction point will be essential.

At the LHC it is also important to consider  background events.
While a full study is deferred to a follow up paper, we determined 
that
the signal over background ratio is enhanced compared to the non
diffractive case because of the good resolution on the dijet mass 
and the
cut on the mass window. Initial studies indicate that the $b 
\bar{b}$ channel
might be interesting to look for Higgs in the diffractive mode.

To summarize, we have shown that the diffractive inclusive
Higgs production
leads to large event rates at the LHC. A Higgs mass
reconstruction with good precision is possible
if  both  protons in the
final state can be tagged 
with roman pot detectors and  if the Pomeron remnants can be 
measured 
in the forward region with sufficient resolution.
This channel and method will  be especially useful in the low mass
Higgs region where the standard methods for Higgs measurements at 
the 
LHC are challenging.

\eject

\vspace{1cm}
 {\bf TABLE}
  \newline
\begin{table}[b] 
\begin{center} 
\begin{tabular}{|c||c|c|c|c|c|c|} 
\hline $M_{Higgs boson}$ &(1)&(2)&(3)&(4)&(5) & (6) \\
\hline\hline  
120 & 3219 & 2043  & 228  & 447  & 48 & 0 \\
150 &2637 &  417  & 48  & 1827  & 222 & 0 \\
200 &1995 &  3   & 0  & 1470 & 522 & 0 \\
300 & 1419 &  0   & 0  & 984  & 438 & 0 \\
375 &  1674 &  0   & 0 & 1047  & 483 & 138  \\
500 & 813 &  0   & 0  & 441  & 213 & 156   \\
700 & 126 &  0 & 0 & 72 & 33 & 18   \\
1000 &  4 & 0 & 0 & 2  & 1 & 0  \\ \hline
\end{tabular} 
\bigskip 
\caption{{\it
Number of Higgs boson events for 10 $fb^{-1}$.}  The first column 
gives the
number of events at the generator level (all decay channels 
included), and the
other columns take into account the roman pot acceptance and
give the number of events for different Higgs boson decay channels 
(2: $b 
\bar{b},$
3: $\tau^+\tau^-,$ 4: $WW,$ 5: $ZZ,$ 6: $t \bar{t},$). 
}  
\end{center} 
\end{table}

{\bf FIGURES}

\bigskip

\input epsf \vsize=8.truecm \hsize=10.truecm \epsfxsize=8.cm{%
\centerline{\epsfbox{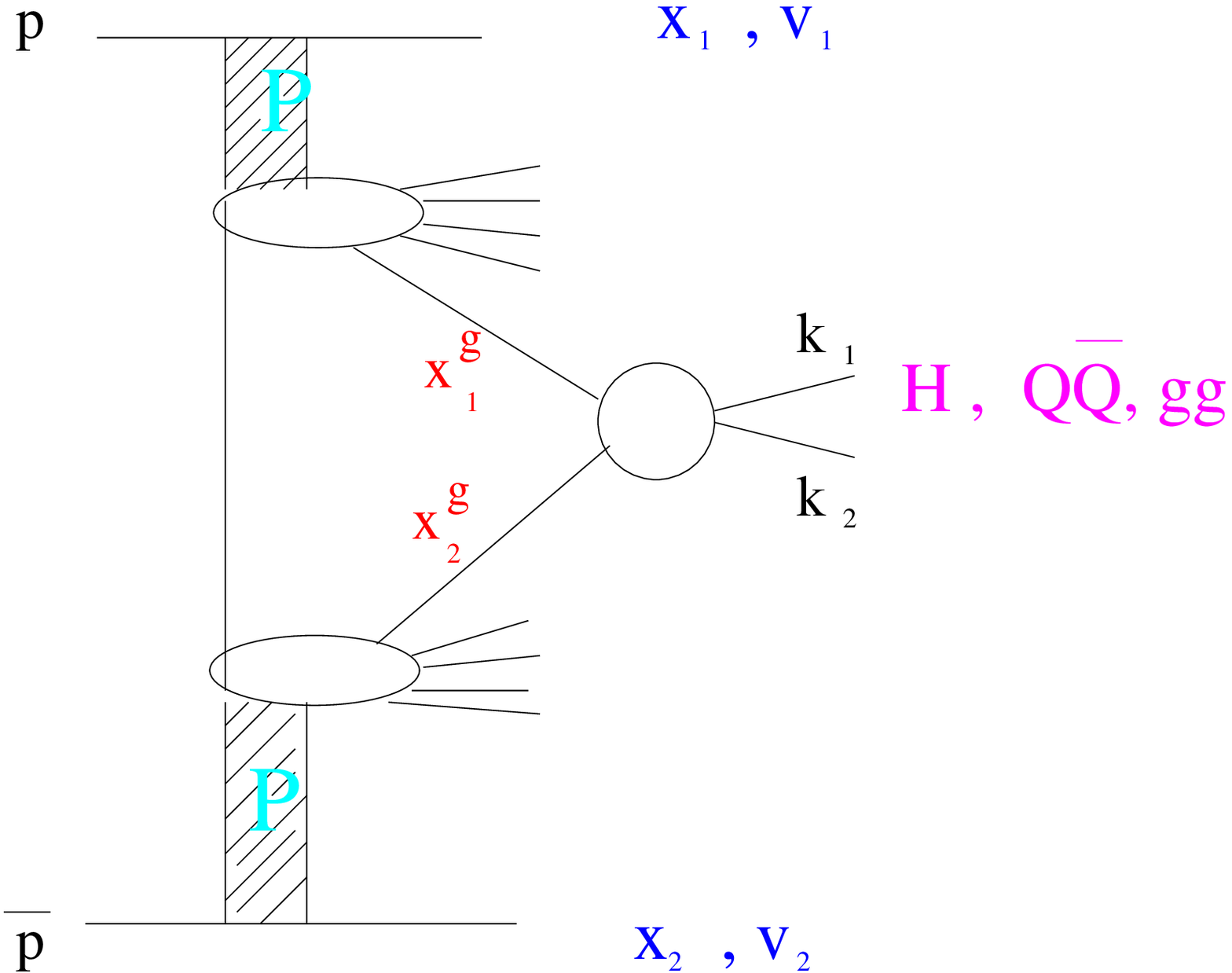}}}\bigskip {\bf Figure 1}

{\it Production scheme.}

\noindent $x_i \equiv 1\!-\!\xi _i,v_i$ are the longitudinal and 
transverse 
2-momenta 
of
the diffracted (anti)proton (see formula (\ref{dinclu}) and text 
for the other 
kinematical 
notations).

\input epsf \vsize=10.truecm \hsize=10.truecm \epsfxsize=10.cm{%
\centerline{\epsfbox{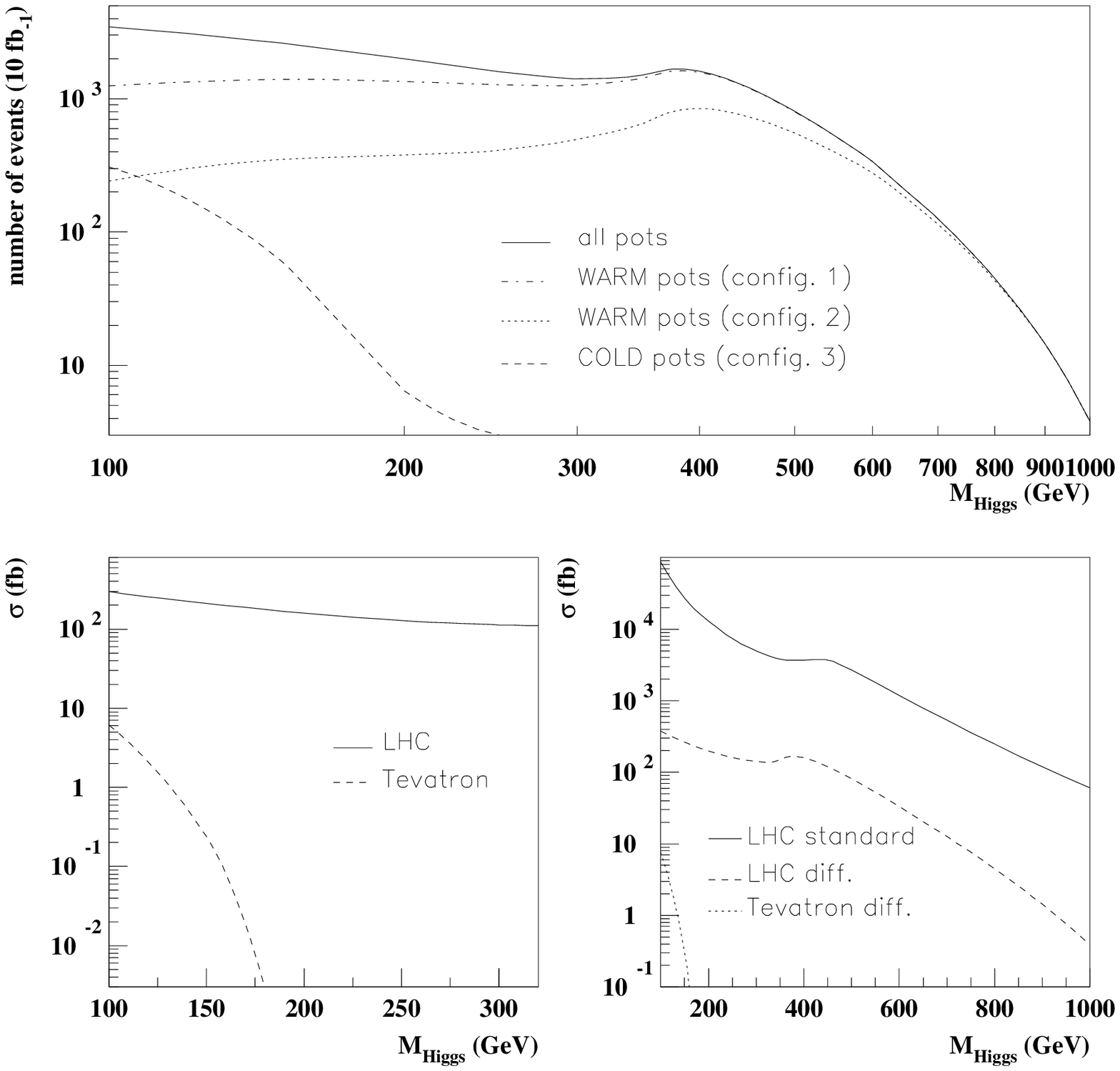}}}\bigskip{\bf Figure 2}

{\it Diffractive Higgs boson production cross section}

\noindent {\it Upper plot}: number of Higgs boson events for 10 
fb$^{-1}$ as a 
function of
$M_H$ obtained for different roman pot configurations (see text).
{\it Bottom plots}: Diffractive Higgs boson production cross 
section as a 
function of 
$M_H$  for the LHC (and the Tevatron). The standard inclusive Higgs 
boson 
production cross
section is also shown for comparison.

\input epsf \vsize=8.truecm \hsize=10.truecm \epsfxsize=8.cm{%
\centerline{\epsfbox{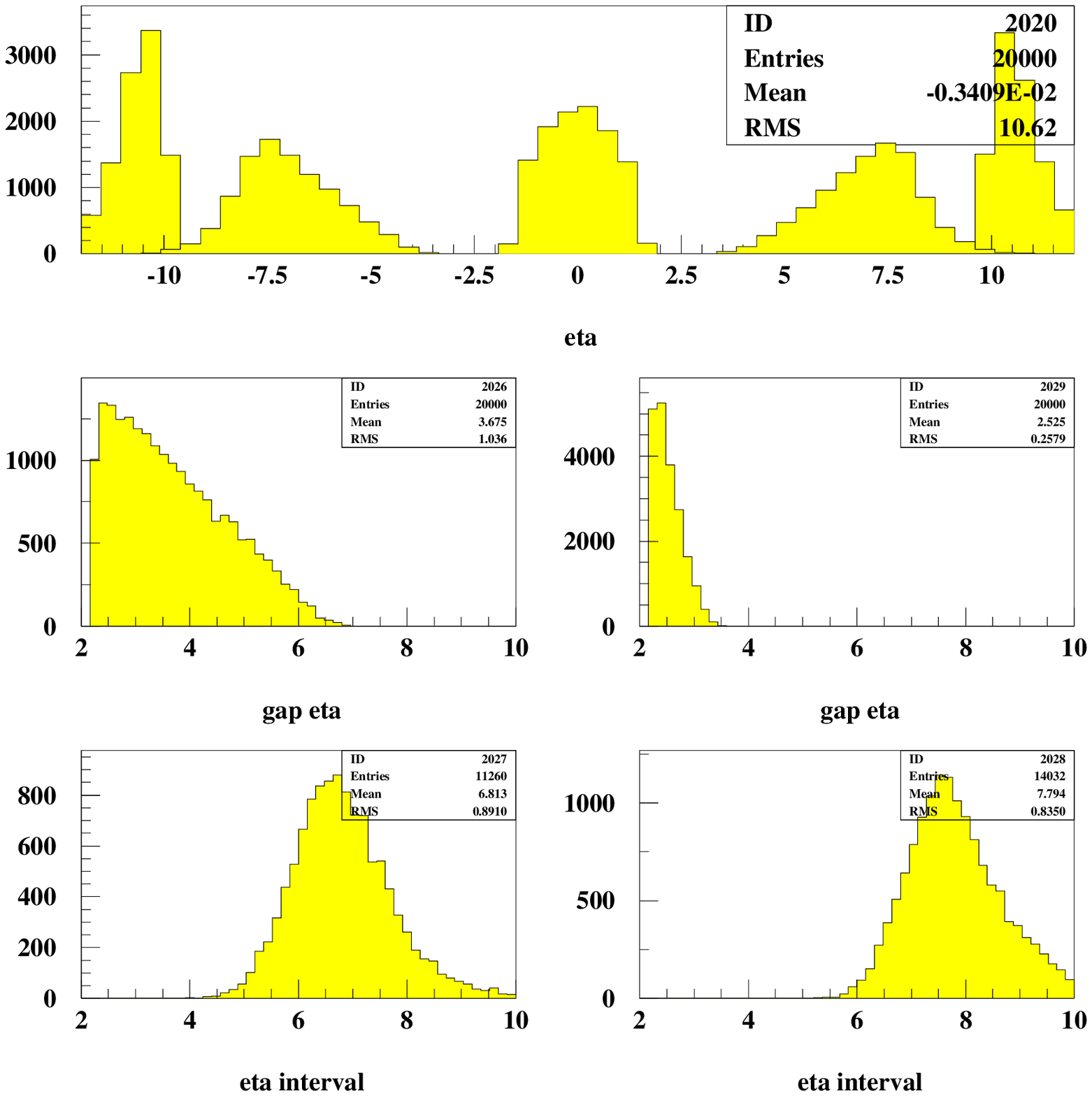}}}\bigskip {\bf Figure 3}

{\it Pseudo-rapidity distributions}

\noindent {\it Upper plot}: pseudo-rapidity distributions of the 
tagged 
protons, Pomeron
remnants  (parton level) and Higgs boson decay products for $M_H = 
120 GeV;$
{\it Medium plots} from left to right: rapidity gap distribution 
between 
the tagged
proton and the Pomeron remnant for two cases $M_H = 120 GeV, M_H = 
700 GeV;$
{\it Bottom plots}: Distance between
the Pomeron remnant (parton level) and the nearest jet from the 
Higgs boson, for 
$M_H = 120 GeV, M_H = 700 GeV.$

\input epsf \vsize=8.truecm \hsize=10.truecm \epsfxsize=8.cm{%
\centerline{\epsfbox{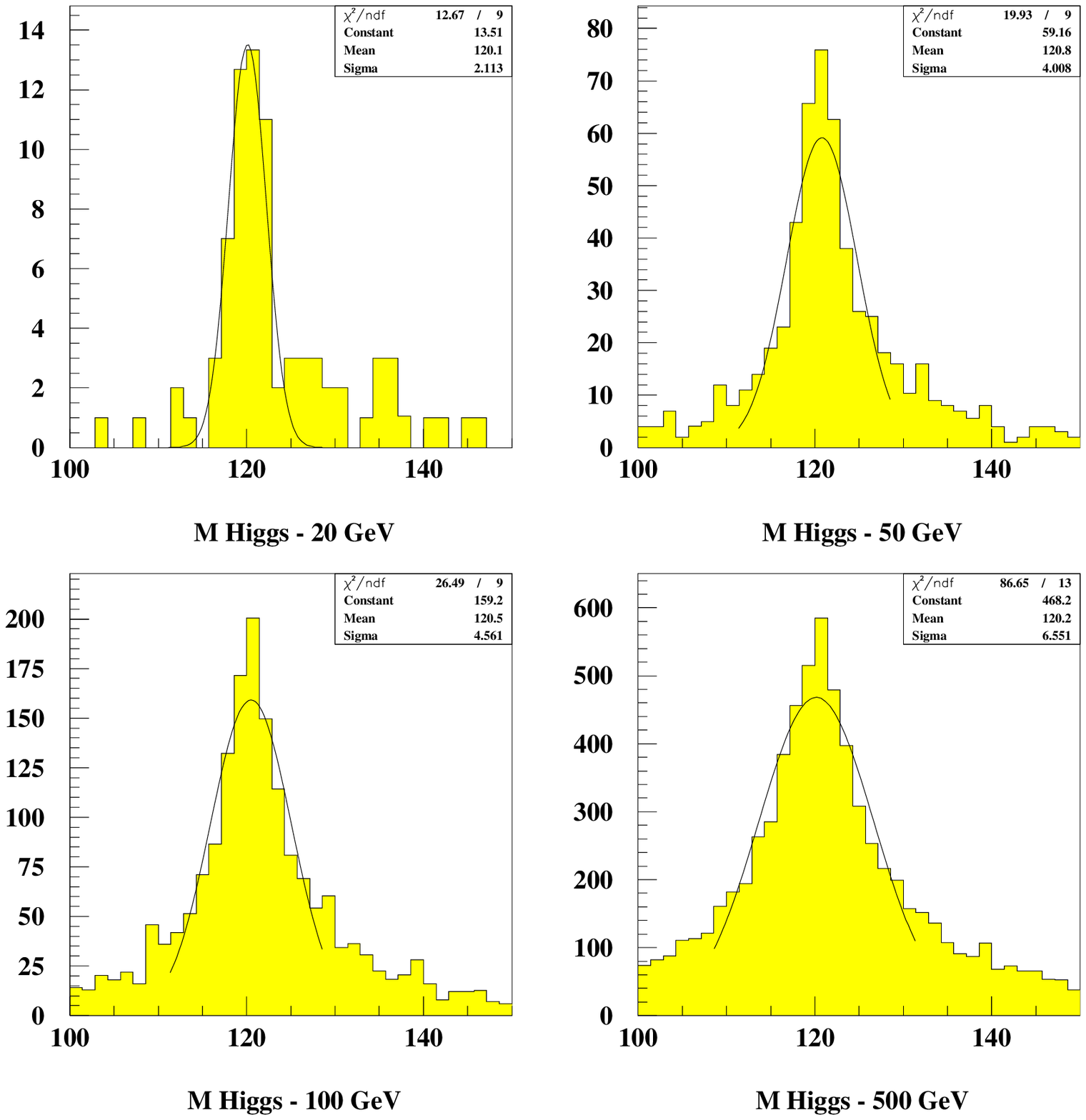}}}\bigskip {\bf Figure 4}

{\it Higgs mass resolution}  

The resolution on the Higgs boson mass is shown after four 
different cuts on the
Pomeron remnant energies at 20, 50, 100 and 500 GeV for a 
luminosity
of 30 fb$^{-1}$.

\end{document}